\documentclass[twocolumn, abstract]{aa} %
\usepackage{graphicx}
\usepackage{natbib}
\bibpunct{(}{)}{;}{a}{}{,}
\usepackage{syntonly}
\usepackage[fleqn]{mathtools}
\usepackage{amssymb}
\usepackage{amsmath}
\usepackage{url}

\usepackage{layouts}
\usepackage{color}
    \definecolor{Blue}{rgb}{0.0,0.0,1.0}
    \definecolor{Red}{rgb}{1.0,0.0,0.0}
    \definecolor{Green}{rgb}{0.0,1.0,0.0}
\newcommand{\be}{\begin{equation}}
\newcommand{\ee}{\end{equation}}
\newcommand{\bea}{\begin{eqnarray}}
\newcommand{\eea}{\end{eqnarray}}
\newcommand{\nn}{\nonumber}
\newcommand{\pp}{\varphi}
\renewcommand\thesection{\arabic{section}}
\begin{document}
%-----------------------------------------------------------------------
\title{A magnetized torus for modeling Sgr~A* \\millimeter images and spectra}

%
%\subtitle{no subtitle yet}
%-----------------------------------------------------------------------
\author{     F. H. Vincent\inst{1}
%-----------------------------------------------------------------------
\and          W. Yan\inst{1}
%-----------------------------------------------------------------------
\and          O. Straub\inst{2}
%-----------------------------------------------------------------------
\and          A. A. Zdziarski\inst{1}
%-----------------------------------------------------------------------
%-----------------------------------------------------------------------
\and          M. A. Abramowicz\inst{1,3,4} }
%-----------------------------------------------------------------------
\institute{
%-----------------------------------------------------------------------
              Nicolaus Copernicus Astronomical Center, ul. Bartycka 18, PL-00-716 Warszawa, Poland
              \\ \email{fvincent@camk.edu.pl}
%-----------------------------------------------------------------------
\and         LUTH, Observatoire de Paris, CNRS, Universit\'e Paris Diderot,
             5 place Jules Janssen, 92190 Meudon, France
             \\ \email{odele.straub@obspm.fr}
\and          Physics Department, Gothenburg University,
              SE-412-96 G{\"o}teborg, Sweden
              \\ \email{marek.abramowicz@physics.gu.se}
%-----------------------------------------------------------------------
\and         Institute of Physics, Faculty of Philosophy and Science, Silesian University in Opava, Bezru{\v c}ovo n{\'a}m. 13,
             CZ-74601 Opava, Czech Republic
%-----------------------------------------------------------------------
}
%-----------------------------------------------------------------------
   \date{Received ; accepted }
%-----------------------------------------------------------------------
%-----------------------------------------------------------------------
\abstract
%............................................................C O N T E X T
%%%   {The observed millimeter properties of the accretion flow in the very vicinity
%%%   of the Galactic center black hole Sgr~A* can be well explained assuming a magnetized
%%%   accretion torus surrounds the compact object. The near-future very long baseline interferometer
%%%   \textit{Event Horizon Telescope} (EHT) will give detailed images of the millimeter surroundings
%%%   of Sgr~A* in the close future.}
%%%
%%%............................O D E L E' S    V E R S I O N    B E L O W
%%%
   {The supermassive black hole, Sagittarius (Sgr)~A*, in the centre of our Galaxy has the largest angular size in the sky among all
   astrophysical black holes. Its shadow, assuming no rotation, spans $\sim 50\,\mu{\rm as}$. Resolving such
   dimensions has long been out of reach for astronomical instruments until a new generation of interferometers being operational
   during this decade. Of particular interest is the \textit{Event Horizon Telescope} (EHT) with resolution $\sim 20\,\mu{\rm as}$ in
   the millimeter-wavelength range $0.87~$mm--$1.3~$mm.}
%..................................................................A I M S
   {We investigate the ability of the fully general relativistic Komissarov (2006) analytical magnetized torus model to account for
   observable constraints at Sgr~A* in the centimeter and millimeter domains. The impact of the magnetic field geometry on
   the observables is also studied.}
%............................................................M E T H O D S
   { We calculate ray-traced centimeter- and millimeter-wavelength synchrotron spectra and images of a 
   magnetized accretion torus surrounding the central black
   hole in Sgr~A*. % for two different geometries of the magnetic field. %For the toroidal one
   We assume stationarity, axial symmetry, constant specific angular momentum and polytropic equation of state. 
%   and small optical depth. 
   A hybrid population of thermal and non-thermal electrons is considered.}
%............................................................R E S U L T S
   {We show that the torus model is capable of reproducing spectral constraints in
   the millimeter domain, and in particular in the observable domain of the EHT. However,
   the torus model is not yet able to fit the centimeter spectrum. $1.3$~mm images at
   high inclinations are in agreement with observable constraints.}
%....................................................C O N C L U S I O N S
   {The ability of the torus model to account for observations of Sgr~A*
   in the millimeter domain is interesting in the perspective of the future EHT.
   Such an analytical model allows very fast computations. It will thus be a suitable test bed
   for investigating large domains of physical parameters, as well as non-black-hole
   compact object candidates and alternative theories of gravity.}
%-----------------------------------------------------------------------
%-----------------------------------------------------------------------
%-----------------------------------------------------------------------
%-----------------------------------------------------------------------
%-----------------------------------------------------------------------
%-----------------------------------------------------------------------
\titlerunning{A magnetized torus for modeling Sgr~A*}
\maketitle
%-----------------------------------------------------------------------
%-----------------------------------------------------------------------
%
%
\section{Introduction}
%
%
%-----------------------------------------------------------------------
%-----------------------------------------------------------------------
A black hole is characterized by its {\it event horizon}, a boundary that causally separates the external universe from its interior and makes the black hole appear black. The reason why there are so far no direct observations of their immediate environment is that astrophysical black holes observed from Earth have a very small apparent angular size in the sky. 
The measurable apparent size of a black hole refers to the size of its {\it shadow}, 
the dark area on the observer's sky cast by the black hole illuminated by nearby radiation. The edge of this shadow
is a very thin ring of light produced by photons approaching the black hole as close as its {\it photon orbit},
i.e. the innermost circular orbit around the black hole.
Photons visiting the region inside the photon orbit may either fall below the event horizon, or may still escape but with such high redshifts that the whole 
area inside the black hole shadow appears dark.
The black hole with largest shadow of all is the supermassive black hole in the centre of our Galaxy, Sagittarius (Sgr) A*. 
Given its mass $M = 4.3 \times 10^6 M_\odot$ and distance $D=8.3$~kpc \citep{ghez08, gillessen09a, gillessen09b}, the angular size of its shadow projected on sky, taking into account the magnification due to gravitational lensing, is only $\approx 50~\mu$as for
a non-rotating black hole. This angular size is a decreasing function of the black hole spin, reaching $\approx 40~\mu$as for a maximally
rotating black hole. Sgr~A* is not the only black hole on sky with such an apparent size: the black hole at the center of the M87 galaxy
reaches an apparent size of $40\,\mu$as (assuming no rotation). However, this article is only dedicated to studying Sgr~A*.

Sgr~A* was first observed in the radio band \citep{balick74}, but its observed emission ranges from radio to X-ray energies. The most remarkable feature of Sgr~A* is its complex variability at all observable wavelengths. The luminosity fluctuations increase with increasing energy, from a factor of a few at radio to a few orders of magnitude in the X-ray band \citep[see e.g.][for a review]{genzel10}. The spectral peak lies in the millimeter radio band and brings forth a peak luminosity of $\lesssim 10^{35}$~erg $\mathrm{s}^{-1}$
(or $\approx 10^{-9}\,L_{\mathrm{Edd}}$). The accretion structure around Sgr~A* is thus extremely dim given its enormous mass. Therefore, adequate disk models describe a radiatively inefficient emitter like an advection dominated accretion flow (ADAF, \citealt{narayan95}, see also the recent review by \citealt{yuan14}).
%or an ion torus \citep[see][and references therein]{straub12}. 
The term {\it advection} means here that a large part of the gravitationally liberated thermal energy is not converted into radiation but carried inward with the ionised, hot accretion flow.

In the millimeter radio range, i.e. at wavelengths corresponding to the spectral peak of Sgr~A*, the {\it Event Horizon Telescope}~\citep[EHT,][]{doeleman09}, operational in 2015-2020, %and the orbital telescope \textit{RadioAstron}~\citep{soko13}, launched in 2011, 
will be able to perform high resolution Very Long Baseline Interferometry (VLBI) observations. 
Like this, images of the close environment of the black hole will be obtained, in particular images of the accretion flow. Recently, the intrinsic emitting region around Sgr~A* was constrained to merely $37~\mu$as by millimeter VLBI~\citep{doeleman08}. Since this is smaller than the shadow size of the presumed black hole, the observed emission should originate from the surrounding accretion flow, seen with a high inclination.
%(from the bright areas left of the silhouette in Figure \ref{fig:images}) in Section 5.

These new observational possibilities on Sgr~A* have stimulated a lot of recent 
research dedicated to modeling the accretion flow surrounding this black hole~\citep[see e.g.][]{goldston05,noble07,chan09,moscibrodzka09,dexter10,shcher12,moscibrodzka13},
as well as works specifically related to the future EHT~\citep[reviewed by][]{broderick14}. The hope is that a detailed knowledge of theoretically
predicted observational appearance of the structure of the accretion flow in Sgr~A* will provide powerful and reliable tools to
test Einstein's general relativity at its strong-field limit. While eventually sophisticated 
general relativistic magnetohydrodynamic (GRMHD) {\it numerical models}~\citep[see e.g.][]{mckinney13} of
Sgr~A* will be used to make a meaningful comparison between theory and observations, in the foreseeable future simple {\it
analytic models} will be invaluable as a secure guide in the vast parameter space that needs to be explored.
In the past 15 years, mainly two such analytic models have been proposed: the radiatively inefficient
accretion flow (\textit{RIAF}) model~\citep{narayan95b,ozel00,yuan03,broderick11} and the \textit{jet} model~\citep{falckemarkoff00}. We propose here
a third analytic model, that we will call the \textit{torus} model.

We have recently constructed an analytic optically thin torus (so-called Polish Doughnut) model of Sgr~A* \citep{straub12}.
It assumed that the magnetic field in Sgr~A* had no global structure, but instead was chaotic i.e. locally isotropic. In this
paper we make the next logical step by considering a model with a globally ordered (toroidal) magnetic field. For this, we
explicitly calculate images and spectra of Sgr~A* in the centimeter and millimeter domains, from $10$ GHz to $10^3$ GHz.
Hereafter, we will refer to the $\approx ~10$~GHz~--~$100$~GHz part of the spectrum as the \textit{centimeter spectrum},
and the to the $\approx ~100$~GHz~--~$10^3$~GHz as the \textit{millimeter spectrum}.
In particular, the millimeter spectrum is relevant for the future EHT observations at the wavelengths
$0.87~$mm and $1.3~$mm ($345$ GHz and $230$ GHz). We use the \citet{komissarov06} analytic model of a magnetized optically thin Polish Doughnut and 
follow all its assumptions. In the Komissarov model, all general relativistic effects, and influence of the (toroidal) magnetic 
field are fully and exactly taken into account. They are calculated from the first principles with no approximation. The presence
of a magnetic field is important in calculations of the synchrotron radiation emissivity, which we also derive following~\citet{wardzinski00}. 
We consider here a torus-shaped, barotropic, and stationary disk with axisymmetry and constant
angular momentum around a Kerr black hole. The disk is fully ionized. These assumptions reflect the basic physics of the real object.
We compute synchrotron spectra emitted by such tori in the range $10$ GHz -- $10^3$ GHz and fit them to
observed data. The spectral fits are compared to the predictions of the RIAF model. 
We compute images at $1.3$ mm and compare their sizes to observable constraints.
We also study the difference in the millimeter spectra and images implied by the magnetic field geometry (chaotic or ordered).
%We modeled In order to compare the predictions of the torus model to those of the RIAF model
%we have also implemented and ray-traced this RIAF model.
%For these two scenarios - chaotic and toroidal magnetic fields - we explicitly calculate images and spectra of Sgr A* relevant 
%for EHT observations, i.e. in the wavelength range $0.87~$mm--$1.3~$mm. 
%Our main goal is to determine whether the Komissarov model can
%account for millimeter observational constraints for Sgr~A*, as well as whether the geometry of the magnetic field (chaotic or toroidal)
%leads to observable differences.

We summarize the basic features of the magnetized torus model and its synchrotron radiation in Sections 2 and 3, respectively. 
Section 4 investigates the ability of the torus model to reproduce spectral observable constraints. Section 5
is dedicated to studying the millimeter images predicted by our model and Section 6 presents conclusions and perspectives.
%--------------------------------------------------------------------
%--------------------------------------------------------------------
%
%
                                  \section{Magnetized accretion torus}
%
%
%--------------------------------------------------------------------
%--------------------------------------------------------------------
In this Section we investigate the observable appearance of an accretion torus for two distinct magnetic field configurations: toroidal
and isotropic (i.e. chaotic). We are also interested in the electron number density distribution, particularly as compared
to the RIAF distribution.
%--------------------------------------------------------------------
%--------------------------------------------------------------------
%
          \subsection{Toroidal magnetic field (the Komissarov model)}
%
%--------------------------------------------------------------------
%--------------------------------------------------------------------
We constructed a magnetized accretion torus at the Galactic centre using the model developed by~\citet{komissarov06}, which
describes analytically a polytropic accretion torus with toroidal magnetic field in the Kerr spacetime.

The fluid 4-velocity is assumed to be
%--------------------------------------------------------------------
\be
%....................................................................
\mathbf{u} = (u^t,0,0,u^\pp)
%....................................................................
\ee
%--------------------------------------------------------------------
using Boyer-Lindquist coordinates. We assume a constant specific angular momentum \be \ell_0 \equiv - u_\pp/u_t. \ee This
quantity is expressed in terms of the dimensionless specific angular momentum \be \lambda = \frac{\ell_0 - \ell_{ms}}{\ell_{mb} -
\ell_{ms}}, \quad 0 \le \lambda \le 1 \ee where $\ell_{ms}$ and $\ell_{mb}$ are the specific angular momentum at the marginally stable and bound orbits
respectively. These assumptions fully determine the 4-velocity.

The gas and magnetic pressures are assumed to follow the polytropic prescription
%--------------------------------------------------------------------
\be
%....................................................................
\label{eq:ppm}
%....................................................................
p = \kappa h^k; \quad p_m = \kappa_m \mathcal{L}^{k-1} h^k
%....................................................................
\ee
%--------------------------------------------------------------------
where $p$ is the gas pressure, $p_m$ is the magnetic pressure, $\kappa$ and $\kappa_m$ are polytropic constants, $k$ is the
polytropic index (assumed identical for gas and magnetic pressures), $h=p+\rho$ is the particle enthalpy where $\rho$ is the gas energy
density, and $\mathcal{L} \equiv g_{t\pp}^2-g_{tt}g_{\pp\pp}$ where $g_{\mu\nu}$ is the Kerr metric.

The conservation of stress-energy leads to
%--------------------------------------------------------------------
\be
%....................................................................
W_s-W+\frac{k}{k-1}\left(\kappa + \kappa_m \mathcal{L}^{k-1}\right)h^{k-1}=0
%....................................................................
\ee
%--------------------------------------------------------------------
where the potential $W=-\mathrm{ln}|u_t|$ is used. We assume that the torus fills its Roche lobe, which fixes the central radius
of the torus and its surface, thus the values of the potential at the centre, $W_c$, and at the surface, $W_s$, of the torus.
This immediately gives 
\be 
h = h_c \left(\omega \frac{\kappa+\kappa_m \mathcal{L}_c^{k-1}}{\kappa+\kappa_m
\mathcal{L}^{k-1}}\right)^{1/(k-1)} 
\ee 
where $h_c$ is the central enthalpy, $\omega = (W-W_s)/(W_c-W_s)$ and $\mathcal{L}_c$ is the value of $\mathcal{L}$ at
the center of the torus.

The polytropic constants $\kappa$ and $\kappa_m$ can be expressed according to
%--------------------------------------------------------------------
\bea
%....................................................................
\kappa &=& h_c^{1-k} (W_c - W_s) \frac{k-1}{k}\frac{\beta_c}{1+\beta_c}, \\
%....................................................................
\nn \kappa_m &=& \frac{\mathcal{L}_c^{1-k}}{\beta_c} \kappa \\
%....................................................................
\nn
%....................................................................
\eea
%--------------------------------------------------------------------
where $\beta_c$ is the central magnetic pressure ratio, $ \beta_c \equiv p_c / p_{m,c}$. The electron number density
%--------------------------------------------------------------------
\be
\label{eq:nbdens}
%....................................................................
n_e = \frac{h-\kappa h^k}{\mu_e m_u}
%....................................................................
\ee
%--------------------------------------------------------------------
is then known analytically, as well as the magnetic pressure.
%, the expression of which is
%\bea
%\label{eq:pm_toro_exp}
%p_m  &=&  h_c \left[\frac{1+\beta_c}{\left(\beta_c+\left(\frac{\mathcal{L}}{\mathcal{L}_c}\right)^{k-1}\right)^{k}}\right]^{1/(k-1)} \frac{k-1}{k} \nn \\ 
%	&& \times (W_c - W_s)   \left(\frac{\mathcal{L}}{\mathcal{L}_c}\right)^{k-1} \omega^{k/(k-1)},  \nn \\
%\eea
%where the first line contains fluid parameters whereas the second line is dictated by 
%spacetime geometry. We note that the magnetic pressure is directly proportional 
%to the central enthalpy, but has a non-trivial dependency on $\beta_c$.

The magnetic pressure has a different expression depending on the magnetic field geometry.
For a chaotic, isotropic field it is $1/3$ of the magnetic energy density, whereas it is equal to
the magnetic energy density for a toroidal field~\citep{hughes91}. Thus, here
\be
%....................................................................
\label{eq:pm_toro}
p_m = \frac{B^2}{8 \pi},   \quad  \mathrm{toroidal \; magnetic \; field}
%....................................................................
\ee
where $B$ is the magnetic field 3-vector magnitude in the fluid frame.

The magnetic field in the Boyer-Lindquist frame is assumed to be toroidal, $\vec{b}=(b^t,0,0,b^\pp)$,
and to be orthogonal to the fluid 4-velocity, $\vec{b}\cdot\vec{u}=0$. 
In the fluid frame, i.e. in an orthonormal tetrad $(\vec{u},\vec{e_i})$, its components are $\vec{b}=(0,0,0,B)$.
The norm of $\vec{b}$ is thus equal to the quantity $B$, the magnitude of the magnetic field 3-vector
in the fluid frame, which is known analytically.
It is then easy to get
%--------------------------------------------------------------------
\bea
%....................................................................
b^\pp &=& \frac{B}{\sqrt{g_{\pp\pp}+2 \ell_0 g_{t\pp}+\ell_0^2 g_{tt}}},\\
%....................................................................
\nn b^t &=& \ell_0 b^\pp. \\
%....................................................................
\nn
%....................................................................
\eea
%--------------------------------------------------------------------
which is fully known analytically. Let us consider one synchrotron photon emitted at a given point of the
torus. Let $\vec{p}$ be the 4-vector tangent to the photon geodesic and $\vec{l}$ be its projection orthogonal to
$\vec{u}$. The angle $\vartheta$ between the magnetic field $\vec{b}$ and the direction of emission is given by $\vec{l} \cdot
\vec{b} = || \vec{l} || \,|| \vec{b} || \,\cos \vartheta$. It is known analytically as well.

We now note that such an accretion torus cannot be made of a perfect gas. If it were, then $ p m_u/(\rho k_B) = T / \mu_e$ where
$T$ is the electron temperature and $k_B$ is the Boltzmann constant. However, it is easy to see that $p/\rho$ is independent of
the central value of the enthalpy $h_c$. Thus the temperature would be independent of $h_c$ as well, and would be purely
determined by the geometry of spacetime, which does not make sense. We will still assume that there exists a relation $T = C p /
\rho$ where $C$ is a constant, but does not take its perfect-gas value. Rather, we choose $T_c$ at the center of the torus and
define the constant $C$ by $T_c = C p_c/\rho_c$. Then
%--------------------------------------------------------------------
\be
%....................................................................
T = T_c \left(\frac{\rho}{\rho_c}\right)^{k-1}
%....................................................................
\ee
%--------------------------------------------------------------------
depends on the choice of $T_c$, and no longer only on spacetime geometry.

%Figure~\ref{fig:nbdens} shows the equatorial plane electron number density, as well as the
%same quantity for the RIAF model for comparison.
%--------------------------------------------------------------------
%--------------------------------------------------------------------
\subsection{Isotropic magnetic field}
%--------------------------------------------------------------------
%--------------------------------------------------------------------
An accretion torus with isotropic (i.e. chaotic) magnetic field has already been studied around Sgr~A* by~\citet{straub12}.
This model is simply the limit of the~\citet{komissarov06} model with $\kappa_m=0$. The section above thus directly
applies to this simpler case. The magnetic field strength is obtained by assuming that the magnetic pressure is everywhere related to
the gas pressure through \be p_m = \frac{1}{\beta} p \ee thus, the $\beta$ parameter is valid in the whole torus, not only at its
center. Then the magnetic field magnitude is known from its link to the magnetic pressure, which for a chaotic
magnetic field is given by~\citep{hughes91}
%--------------------------------------------------------------------
\be
%....................................................................
\label{eq:pm_iso}
p_m = \frac{B^2}{24 \pi},   \quad  \mathrm{chaotic \; magnetic \; field.}
%....................................................................
\ee
%--------------------------------------------------------------------

%At this point, all thermodynamical quantities are known analytically,
%together with the magnetic field norm in the comoving frame and
%the angle between the magnetic field and the direction of emission.
%Synchrotron emission can then be computed.

\subsection{Torus cross-section and density distribution}

This Section aims at giving an overview of how the accretion torus properties depend
on its parameters, and in particular to compare its density distribution to the RIAF model.

\begin{figure*}[htbp]
\centering
\includegraphics[width=0.4\hsize]{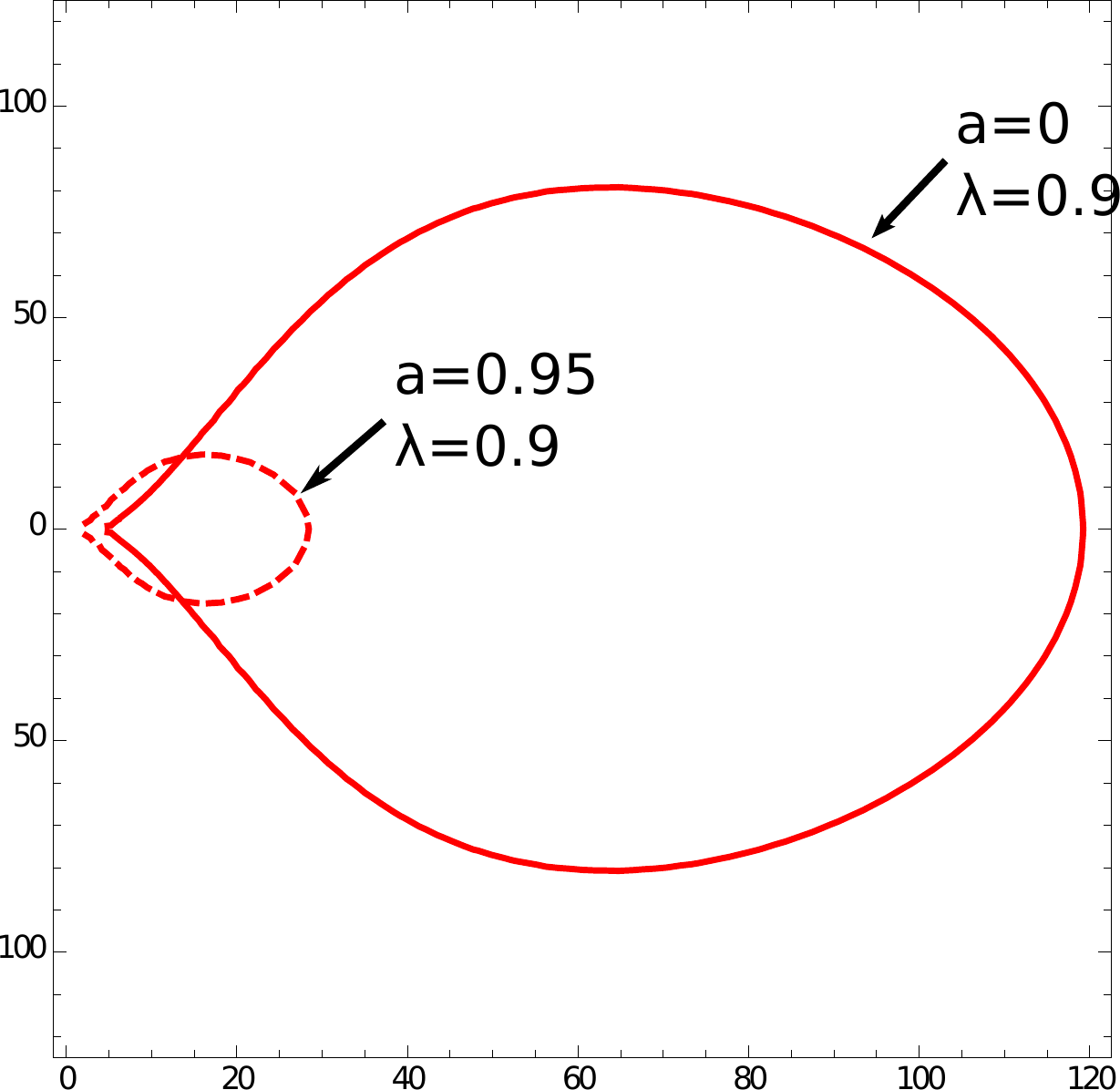}
\caption{Cross-section of the torus surface for a fixed angular momentum $\lambda=0.9$ when varying the 
spin parameter from $a=0$ (solid line) to $a=0.95$ (dashed line). 
As the spin increases, the torus becomes smaller and its inner radius decreases.
The axes are in units of the gravitational radius $GM/c^2$.} \label{fig:cross}
\end{figure*}

The cross-section of the torus is dictated by the spin value and angular momentum
$\lambda$ parameter. Fig.~\ref{fig:cross} shows the cross-section of a torus for
$\lambda=0.9$ and spin $a=0$ and $a=0.95$. The outer radius of the torus strongly decreases
when increasing spin, from $r_{\mathrm{out}}=120\,r_g$ to $r_{\mathrm{out}}=30\,r_g$ in the
case illustrated. The inner radius of the torus (the Roche lobe overflowing point) decreases
when spin increases, from $r_{\mathrm{in}}=4\,r_g$ at spin $0$ to $r_{\mathrm{in}}=1.5\,r_g$
at spin $0.95$. An important point is that the central radius, where all physical quantities
(density, temperature, magnetic field) reach their maximum, is very different from the geometrical
center. In our example case, it goes from $r_{\mathrm{center}}=10\,r_g$ at spin $0$
to $r_{\mathrm{center}}=3\,r_g$ at spin $0.95$. In all cases, the central radius, i.e. the location
of the maximum of emission, is at only a few $r_g$. In the optically thin part of the spectrum (i.e.,
the millimeter spectrum essentially), only the very central part of the torus will contribute
to the flux, no matter what the value of $\lambda$ is.

\begin{figure*}[htbp]
\centering
\includegraphics[width=0.4\hsize]{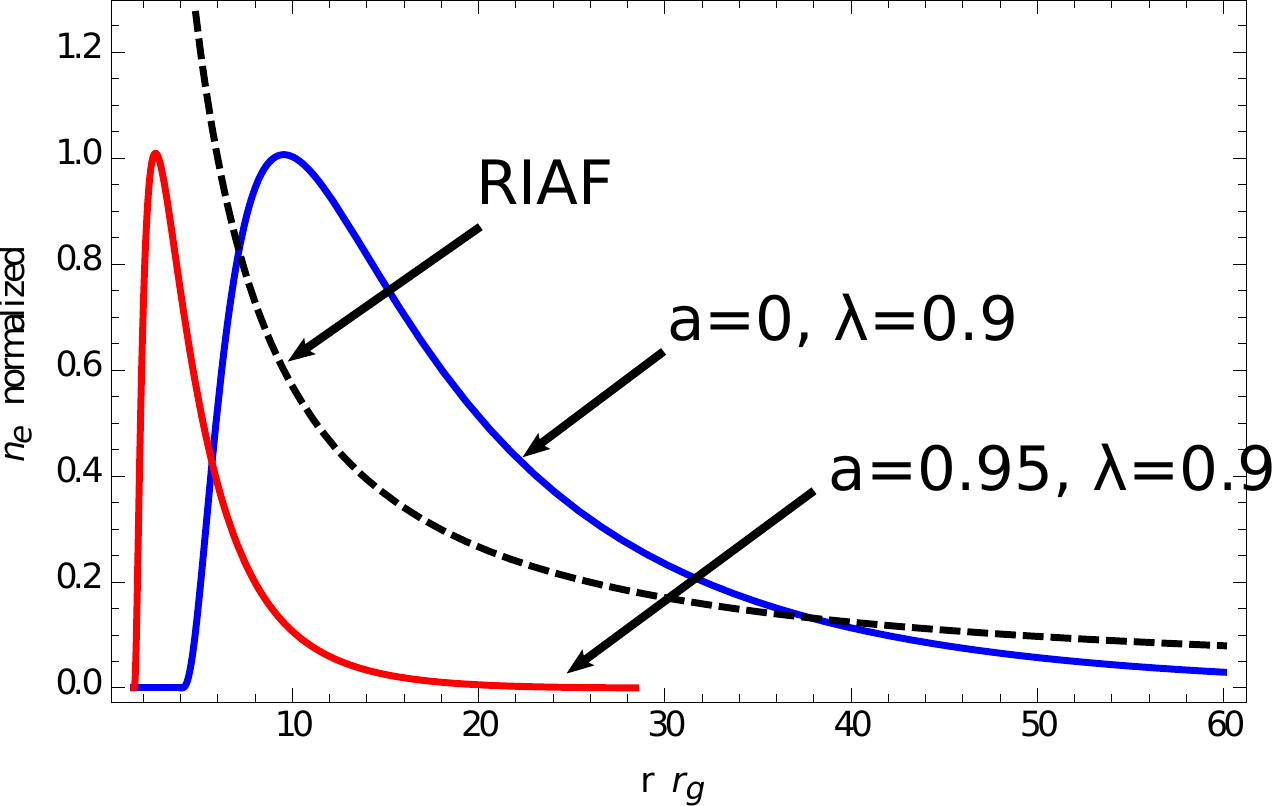}
\caption{Normalized distribution of the electron number density for the
torus model with spin $0$ and $\lambda=0.9$ (solid blue) or
spin $0.95$ and $\lambda=0.9$ (solid red). For comparison, the RIAF
power-law distribution is plotted in dashed black~\citep[from][]{broderick11}.} \label{fig:nbdens}
\end{figure*}

The electron number density distribution (Eq.~\ref{eq:nbdens}) is a rather
non-trivial function of radius and angle. Fig.~\ref{fig:nbdens} illustrates its profile 
in the equatorial plane as a function of radius for the
same cases illustrated in Fig.~\ref{fig:cross}, and compares it to the RIAF
number density distribution. The RIAF distribution is taken from \citet{broderick11},
Eq.~2: it is a simple power law with slope $-1.1$.
The torus model distribution is clearly not a power law, it is concentrated
around the central radius of the torus, decreasing much more rapidly than
the RIAF power law for bigger radii. As a consequence, the distant regions of
the torus model will contribute less than the distant regions of the RIAF model.
This point will be important later on. 

\section{Synchrotron radiation}
%
%
%--------------------------------------------------------------------
%--------------------------------------------------------------------
In this Section, we express the synchrotron emission and absorption
coefficients for a directional magnetic field, and
for either a thermal or non-thermal population of electrons.
Radiative transfer for an angle-averaged magnetic field is
simply obtained by angle-averaging the directional expressions.
%--------------------------------------------------------------------
%--------------------------------------------------------------------
%
\subsection{Thermal directional synchrotron radiation}
%
%--------------------------------------------------------------------
%--------------------------------------------------------------------
\citet{wardzinski00} show that for a mildly relativistic Maxwellian electron distribution,
%--------------------------------------------------------------------
\begin{equation}
%....................................................................
n_e(\gamma)=\frac{n_e}{\theta_e}\frac{\gamma(\gamma^2 - 1)^{1/2}}{K_2(1/\theta_e)}\exp[-\frac{\gamma}{\theta_e}],
%...................................................................
\end{equation}
%--------------------------------------------------------------------
where $\theta_e = k_B T / (m_e c^2)$, $m_e$ being the electron mass, $\gamma=(1-v^2/c^2)^{-1/2}$ is the Lorenz factor and
$K_2$ is a modified Bessel function, the emission coefficient is
%--------------------------------------------------------------------
\begin{equation}
%....................................................................
\begin{split}
%....................................................................
j_\nu^{\mathrm{dir,th}}=&\frac{\pi e^2}{2c}(\nu\nu_0)^{1/2}\mathcal X(\gamma_0) n_e(\gamma_0)\left(1 +2\frac{\cot^2\vartheta}{\gamma_0^2}\right)\\
%....................................................................
&\times\left[1 - \left(1 - \gamma_0^{-2}\right) \cos^2\vartheta\right]^{1/4}\mathcal Z(\vartheta,\gamma_0)
%....................................................................
\end{split}
%....................................................................
\end{equation}
%--------------------------------------------------------------------
where $\nu_0\equiv eB / (2\pi m_e c)$ is the cyclotron frequency. 
%The superscript dir means that this emission coefficient depends
%on the angle between the magnetic field and the direction of emission, no angle averaging has been performed. 
Then,
%--------------------------------------------------------------------
\begin{equation}
%....................................................................
\gamma_0=
%....................................................................
\begin{cases}
%....................................................................
\displaystyle{\left[1 + \frac{2\nu\theta_e}{\nu_0} \left(1 + \frac{9 \nu\theta_e\sin^2\vartheta}{2 \nu_0}\right)^{-\frac 1 3}\right]^{\frac 1 2}}
&  \quad\theta_e \lesssim 0.08 \\ \\
%....................................................................
\displaystyle{\left[1 + \left(\frac{4\nu\theta_e}{3 \nu_0\sin\vartheta}\right)^{\frac 2 3}\right]^{\frac 1 2}}  &  \quad\theta_e \gtrsim 0.08
%....................................................................
\end{cases}
%....................................................................
\end{equation}
%--------------------------------------------------------------------
is the Lorenz factor of those thermal electrons that contribute most to the emission at $\nu$, and
%--------------------------------------------------------------------
\begin{equation}
\mathcal X(\gamma)=
%...................................................................
\begin{cases}
%...................................................................
\displaystyle{\left[\frac{2\theta_e(\gamma^2 - 1)}{\gamma(3 \gamma^2 - 1)}\right]^{1/2}},  &  \quad\theta_e \lesssim 0.08\\
\\
%...................................................................
\displaystyle{\left(\frac{2\theta_e}{3\gamma}\right)^{1/2}},  &  \quad\theta_e \gtrsim 0.08
%...................................................................
\end{cases}
%...................................................................
\end{equation}
%-------------------------------------------------------------------
%
%-------------------------------------------------------------------
\begin{equation}
%...................................................................
\begin{split}
%...................................................................
&t \equiv (\gamma^2 - 1)^{\frac 1 2}\sin\vartheta,    \quad     n\equiv \frac{\nu(1 + t^2)}{\nu_0\gamma},\\
%...................................................................
&\mathcal Z(\vartheta,\gamma)=\left\{\frac{t\exp[(1 + t^2)^{-\frac 1 2}]}{1 + (1 + t^2)^{\frac 1 2}}\right\}^{2n}.
%...................................................................
\end{split}
%...................................................................
\end{equation}
%-------------------------------------------------------------------

The absorption coefficient is simply given by Kirchhoff's law: $\alpha_\nu^{\mathrm{dir,th}} = j_\nu^{\mathrm{dir,th}} / B_\nu$
where $B_\nu$ is the Planck blackbody function.

We note that our ray-tracing code allows to naturally take into account
synchrotron self-absorption. The radiative transfer equation is simply solved
along null geodesics until the torus becomes optically thick.

\subsection{Non-thermal directional synchrotron radiation}
\label{sec:PLelec}

Motivated by the RIAF model which is able to fit the whole radio (centimeter and millimeter) spectrum of Sgr~A*
using a population of non-thermal power-law electrons, we implemented such
a population in our torus model.

We consider a population of electrons with a power-law energy spectrum
\be
f(\gamma) = n_e^{\mathrm{pl}} (p-1) \gamma^{-p}
\ee
where $n_e^{\mathrm{pl}}$ is the power-law electron number density.
Following~\citet{ozel00} we assume that the energy of the non-thermal
electrons is equal a small fraction $\delta$ of the
energy of the thermal electrons. This leads to
\be
n_e^{\mathrm{pl}} = \delta a(\theta_e) \theta_e (p-2) n_e
\ee
where $a(\theta_e)$ is an order-unity function of temperature
given by~\citet{ozel00}, Eq.~6.

The power-law emission and absorption coefficients are then given by~\citep{petrosian83}
\bea
j_\nu^{\mathrm{dir,pl}} &=& \frac{\sqrt{3} \pi e^2 \nu_0 \sin \vartheta}{2c} n_e^{\mathrm{pl}} (p-1)  \\ \nonumber
&& \times \left(\frac{3 \nu_0 (p+1) \sin \vartheta}{4\nu}\right)^{(p-1)/2} \mathrm{exp}\left(-\frac{p+1}{2}\right), \\ \nonumber
\alpha_\nu^{\mathrm{dir,pl}} &=& \frac{\sqrt{3} \pi e^2 \nu_0 \sin \vartheta}{2 c \,m_e \nu^2} n_e^{\mathrm{pl}} (p-1) (p+2)  \\ \nonumber
&& \times \left(\frac{3 \nu_0 (p+2) \sin \vartheta}{4\nu}\right)^{p/2} \mathrm{exp}\left(-\frac{p+2}{2}\right). \\ \nonumber
\eea
We note here a typo in~\citet{ozel00} and~\citet{yuan03}: the frequency exponent of their
absorption coefficient is not correct.

\section{Torus model spectra}
%-------------------------------------------------------------------
%-------------------------------------------------------------------

We use the open-source\footnote{Freely available at the URL \url{http://gyoto.obspm.fr}} ray-tracing code
\texttt{GYOTO}~\citep{vincent11} to compute images and spectra of the magnetized accretion torus. 
We have studied the impact of the models' parameters on the observables in order to produce
simulations that are in agreement with current constraints. 
Restricting ourselves to the centimeter and millimeter domains,
we want our model to be able to reproduce the observed spectrum of Sgr~A* (this Section). The corresponding
images should predict a correct size of the emitting zone which was constrained at $1\sigma$ to $37^{+5.3}_{-3.3}\, \mu$as 
at $1.3~$mm (intrinsic size) or $43^{+4.7}_{-2.7}\, \mu$as (scatter broadened) by~\citet[][next Section]{doeleman08}. 

In order to allow a direct comparison with the angle-averaged RIAF model, all torus
model simulations are performed with angle-averaging unless otherwise stated.

\subsection{Fitting the millimeter spectrum}

The RIAF model is able to fit equally well any values of spin and inclination:
these two parameters are not constrained by spectral data only~\citep{broderick11}.
We obtain the same behavior with our torus model: spin and inclination are not
constrained from the millimeter spectral data only. 

Our torus model is parameterized by a set of $7$ parameters listed in Tab.~\ref{tab:param}.
Here, we have always kept constant the plasma $\beta$ parameter, to $\beta=10$,
as well as the polytropic index $k=5/3$.
For a given pair of spin and inclination, there are thus $3$ parameters to fit: 
the torus constant angular momentum $\lambda$, its central temperature $T_c$
and central density $n_c$.
We have performed a standard minimum $\chi^2$ analysis, minimizing the
distance from model to data on a grid of parameters. 
\begin{table}[htbp!]
\centering \caption{Torus model parameters.
The value of the parameter is given if it is not fitted. The central density is defined by
$n_c~=~h_c / (m_\mathrm{u} c^2)$, where $m_{\mathrm{u}}$ is the atomic mass unit. }
\begin{tabular}{l{c}r}
parameter                  &             & value                                \\
\hline
spin                       & $a$         &                                       \\
inclination                & $i$         &                \\
angular momentum           & $\lambda$   &                                   \\
gas/magnetic pressure ratio              & $\beta_c$, $\beta$ & $10$                 \\
polytropic index           & $k$           & $5/3$                                     \\
central density ($ \mathrm{cm}^{-3}$)           & $n_c$       &                                \\
central electron temperature (K)            & $T_c$   &       \\
\end{tabular}
\label{tab:param}
\end{table}
We have been considering
a grid of $(\lambda,T_c,n_c)$ with $\lambda$ varying from $0.3$ to $0.9$,
$T_c$ from $7\times10^{10}$~K to $3\times10^{12}$~K, and
$n_c$ from $3\times10^6 \, \mathrm{cm}^{-3}$ to $10^7 \, \mathrm{cm}^{-3}$. 
We have considered $30\times30\times10$
values of $(\lambda,T_c,n_c)$ within these boundaries.
We have investigated only extreme values of the spin parameter, 
$a=0$ and $a=0.95$, considering three values of inclination for
both cases ($i=5^\circ$, $i=45^\circ$, $i=85^\circ$). Inclination is defined
from the axis of rotation of the black hole to the line of sight: $i=5^\circ$
means nearly face-on view.

\begin{figure*}[htbp]
\centering
\includegraphics[width=0.4\hsize]{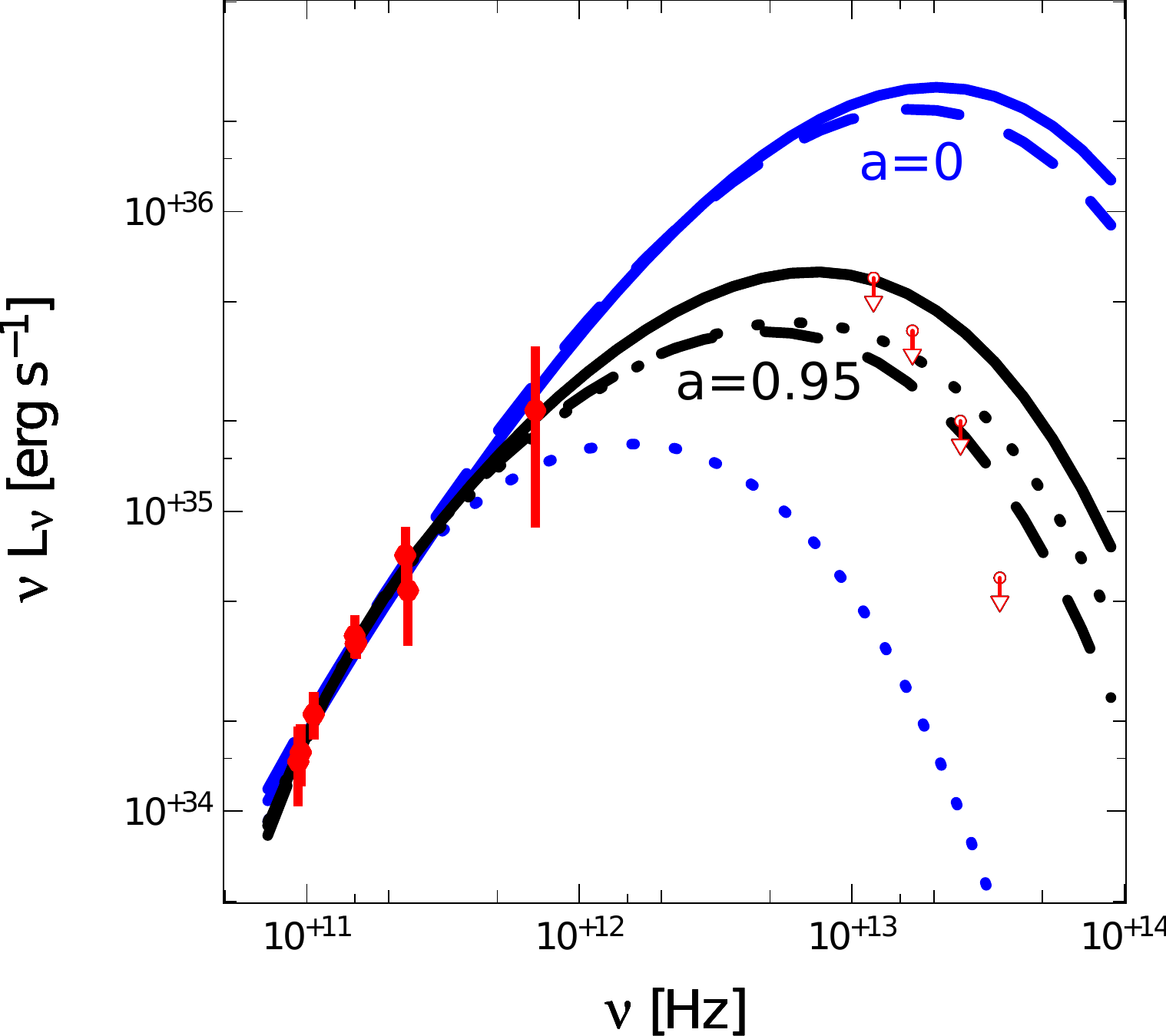}
\caption{Best-fitting spectra obtained for the parameters listed in Tab.~\ref{tab:parambf}.
Blue curves are for spin $0$, black for spin $0.95$. Solid line is for $i=5^\circ$,
dashed for $i=45^\circ$, and dotted for $i=85^\circ$. Note that the fitting
is done only on the millimeter error bars. The infrared upper limits are
given for completeness, but are not taken into account in the fit.} \label{fig:specbf}
\end{figure*}

Tab.~\ref{tab:parambf} gives the best-fit parameters for these configurations.
It shows that all fits lead to $\chi^2_{\mathrm{red}} \approx 0.2$--$0.4$, 
thus to very good fits. Such low values of $\chi^2_{\mathrm{red}}$ are
obtained because of the large error of the $690$~GHz data (see Fig.~\ref{fig:specbf}).
This large error reflects the variability of the source
rather than instrumental precision~\citep[see][]{marrone06}. All values of
spin and inclination can thus be fitted with very good accuracy: the spectral millimeter data
alone are not sufficient to constrain these parameters. 

It is interesting to note that all best-fit physical values are rather close to
each other, with $r_{\mathrm{out}} \approx 15$, $n_c \approx 5\times10^6\,\mathrm{cm}^{-3}$,
$T_c \approx 5\times10^{11}$~K. These values are in good agreement with 
other models of Sgr~A* accretion flow, be they analytic or numeric~\citep[see e.g.][]{broderick06,moscibrodzka09}.
A particularly striking point
is the clear preference for very compact configurations, with the outer boundary
of the torus being in all cases at $10$--$20\,r_g$, translating
to approximately $70\,\mu$as from the black hole position. This makes a big difference
with respect to the RIAF model, for which the distribution of electrons is
extended to regions far away from the black hole.
\begin{table}[htbp!]
\centering \caption{Torus model best-fit parameters for $2$ extreme
values of spin and a range of inclinations. The third column gives the
outer radius corresponding to the best fitting $\lambda$ (this outer radius
strongly depends on spin for a given $\lambda$).}
\begin{tabular}{cccccc}
($a$,$i$)                  &      $\lambda$  & $r_{\mathrm{out}} \, (r_g)$       & $n_c \,(\mathrm{cm}^{-3})$ &  $T_c$ (K)     &         $\chi^2_{\mathrm{red}}$                 \\
\hline
($0$,$5^\circ$)               & $0.35$  &    $15$      &  $7.7\times10^6$       &     $8.7\times10^{11}$    &    $0.37$                  \\
($0$,$45^\circ$)            & $0.33$  &    $15$         &    $8.4\times10^6$     & $7.5\times10^{11}$      & $0.37$     \\
($0$,$85^\circ$)            & $0.33$  &    $15$         &    $5.6\times10^6$     & $2.3\times10^{11}$      & $0.25$     \\
($0.95$,$5^\circ$)          & $0.75$  &    $11$         &    $1\times10^7$     & $4.2\times10^{11}$      & $0.21$     \\
($0.95$,$45^\circ$)         & $0.79$  &    $13$         &    $7\times10^6$      &  $3.1\times10^{11}$   & $0.21$     \\
($0.95$,$85^\circ$)         & $0.85$  &    $20$         &    $3.5\times10^6$     & $3.1\times10^{11}$      & $0.21$     \\
\end{tabular}
\label{tab:parambf}
\end{table}

Fig.~\ref{fig:specbf} shows the $6$ spectra associated to the best fits in Tab.~\ref{tab:parambf}.
It is interesting to note that the thermal bump of the spectrum is always obtained at
rather high frequency (between $10^{12}$~Hz and $5\times10^{13}$~Hz). For completeness, Fig.~\ref{fig:specbf}
also shows the near-infrared upper limits: it is likely that some spin/inclination configuration
might be excluded when fitting from radio to infrared data. {However, here, the fit
is only done on millimeter data, which means on the data points in Fig.~\ref{fig:specbf}
with frequency between $\approx 10^{11}$~Hz and $10^{12}$~Hz. The infrared upper limits
are not taken into account in the fitting procedure.}

\subsection{Robustness of the fit results}

We have analyzed in more detail the distribution of $\chi^2$ for the two
cases in Tab.~\ref{tab:parambf} corresponding to $i=85^{\circ}$. This choice
was dictated by the fact that higher inclination is favored by the imaging
constraints (see next Section).

The first important result is that the size of the torus is indeed firmly
constrained to values of $r_{\mathrm{out}} \lesssim 20 \,r_g$. At spin
$0$, all $\chi^2_{\mathrm{red}}$ are above $1$ for $\lambda > 0.63$,
and all $0.5 \le \lambda \le 0.63$ lead to spectra distant by more than
$1\sigma$ from the $690$~GHz data, which has a very large error bar
due to the source variability. We conclude that we can safely constrain
the angular momentum to $\lambda < 0.5$, equivalent to $r_{\mathrm{out}} < 20$.
We obtain a comparable constraint from the $a=0.95$ fit. The torus model
is thus imposing a compact configuration.

\begin{figure*}[htbp]
\centering
\includegraphics[width=0.8\hsize]{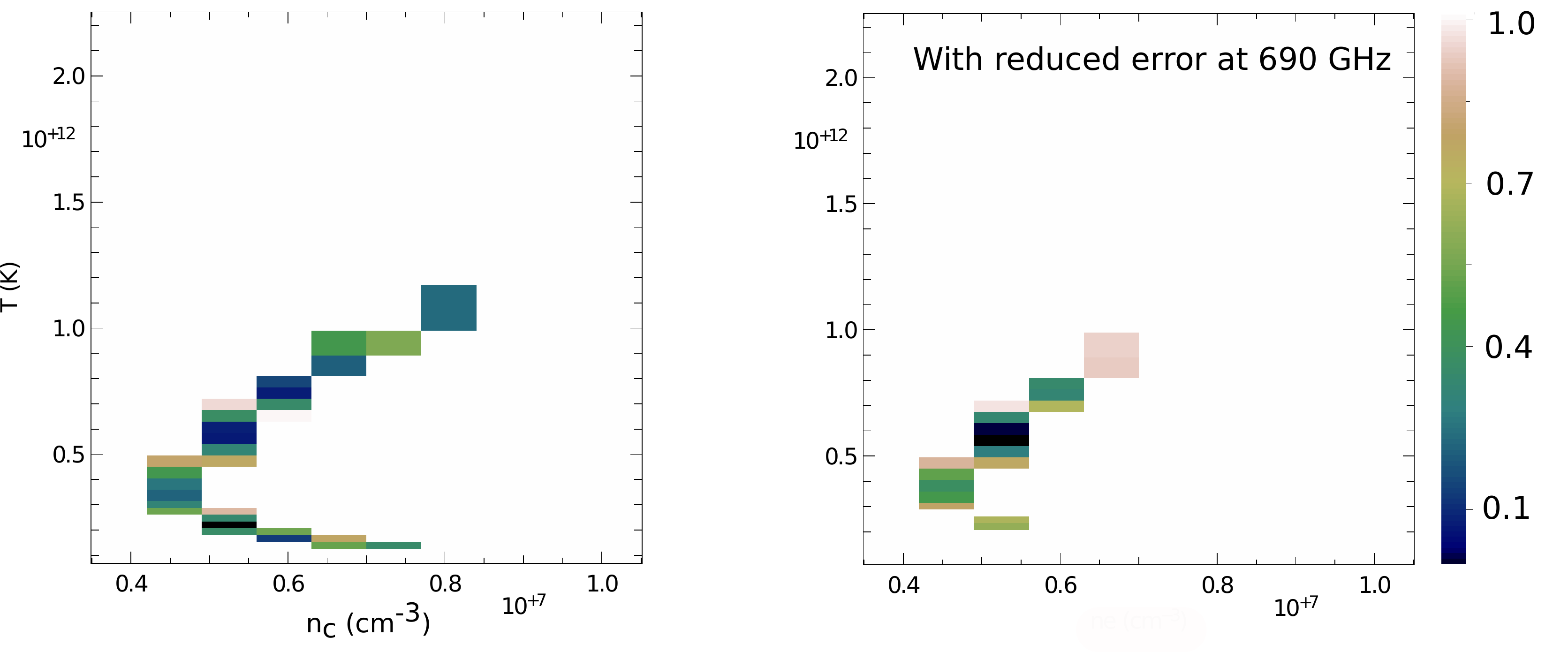}
\caption{$\chi^2_{\mathrm{red}}$ distribution in the $(n_c,T_c)$ plane, for the $a=0$, $i=85^\circ$ fit. All values of
$\chi^2_{\mathrm{red}}$ bigger than $1$ have been put to a high value and
appear in white color in this plane to focus on the relevant configurations
with $\chi^2_{\mathrm{red}}<1$. \textbf{Left}: using
the error bar at $690$~Hz used in the fitting routine. \textbf{Right}: reducing
the $690$~GHz error by a factor of $3$: the region with $\chi^2_{\mathrm{red}}<1$
immediately shrinks. We note that the best-fit temperature changes, while the
best-fit density remains the same.} \label{fig:chi2plane}
\end{figure*}

Fig.~\ref{fig:chi2plane} shows the $\chi^2_{\mathrm{red}}$ 2D distribution
in the $(n_c,T_c)$ plane for the best-fitting value $\lambda=0.33$ at
spin $0$ and inclination $85^\circ$. From the left panel, the central
density is rather well constrained (up to a factor of $2$), but the central
temperature is allowed to vary by a factor of $\approx 10$, always
giving $\chi^2_{\mathrm{red}}<1$. This freedom is only due to the
large error bar at $690$~GHz. The right panel of Fig.~\ref{fig:chi2plane}
shows the same distribution, but this error has been reduced by a 
factor of $3$. The central temperature is immediately much more
constrained. However, the best-fit value of $T_c$ with a reduced
error bar at $690$~GHz differs from its counterpart with the unmodified
error bar (while the central density remains unchanged).
The central temperature changes from $T_c = 2.3\times10^{11}$~K
to $T_c = 5.5\times10^{11}$~K. A very similar analysis could be made
with spin $0.95$ fit.

We conclude from this analysis that the fit results presented
in Tab.~\ref{tab:parambf} are robust as far as the angular momentum
(or torus outer radius) and central density are concerned. The
central temperature is less robust %(to within a factor of $\approx2$)
due to the large error bar at $690$~GHz.

\subsection{Impact of magnetic field geometry on the millimeter spectrum}

\begin{figure*}[htbp]
\centering
\includegraphics[width=0.4\hsize]{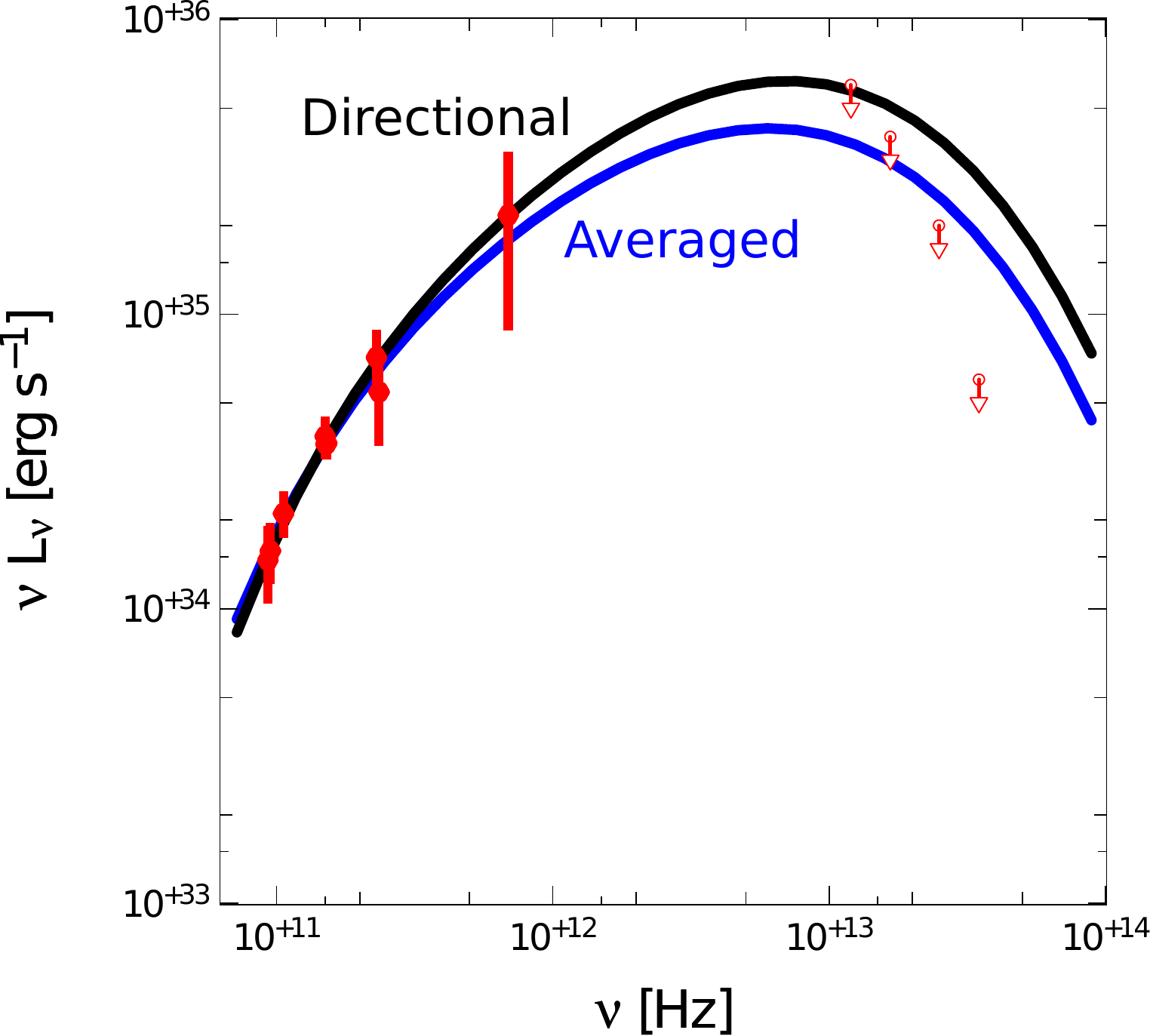}
\caption{Comparison of the best-fit $a=0.95$, $i=85^\circ$ spectra 
for an angle-averaged magnetic field (blue) and a directional toroidal
magnetic field (black). Spectral data are not enough to put constraints
on the magnetic field geometry. {The infrared upper limits are
given for completeness, but are not taken into account in the fit.}} \label{fig:specAvg}
\end{figure*}

One of our interest in this article is to investigate possible observable
differences due to the magnetic field geometry. Fig.~\ref{fig:specAvg}
shows one of the best-fit spectra ($a=0.95$, $i=85^\circ$), obtained
for an angle-averaged configuration, together with the best-fit spectrum
associated to a directional magnetic field. This directional best fit is found 
for the parameters $\lambda=0.85,n_c=6.3\times10^6~\mathrm{cm}^{-3}, T_c=3.6\times10^{11}~K$. 
The small difference
between these directional parameters and their angle-averaged
counterparts is mainly due to the difference in the relation between
magnetic pressure and magnetic field magnitude (see Eqs.~\ref{eq:pm_toro}
and~\ref{eq:pm_iso}).

The difference between the best-fit spectra is weak and it is not very
surprising to conclude that spectral data
will not be helpfull to constrain the magnetic field geometry.

\subsection{No fit for the centimeter part of the spectrum}

So far, we have been focused only on the millimeter part of
Sgr~A* spectrum, which is likely to be accounted for by thermal
synchrotron. Lower radio frequencies are not fitted by a pure
thermal population as the low-frequency slope of the thermal
spectrum is too hard to fit the data. The RIAF model is able to
fit this centimeter part of the spectrum by invoking a small proportion
of non-thermal, power-law electrons~\citep{ozel00}. We have
thus considered such a population in our model and tried to
fit for the centimeter data.

\begin{figure*}[htbp]
\centering
\includegraphics[width=0.4\hsize]{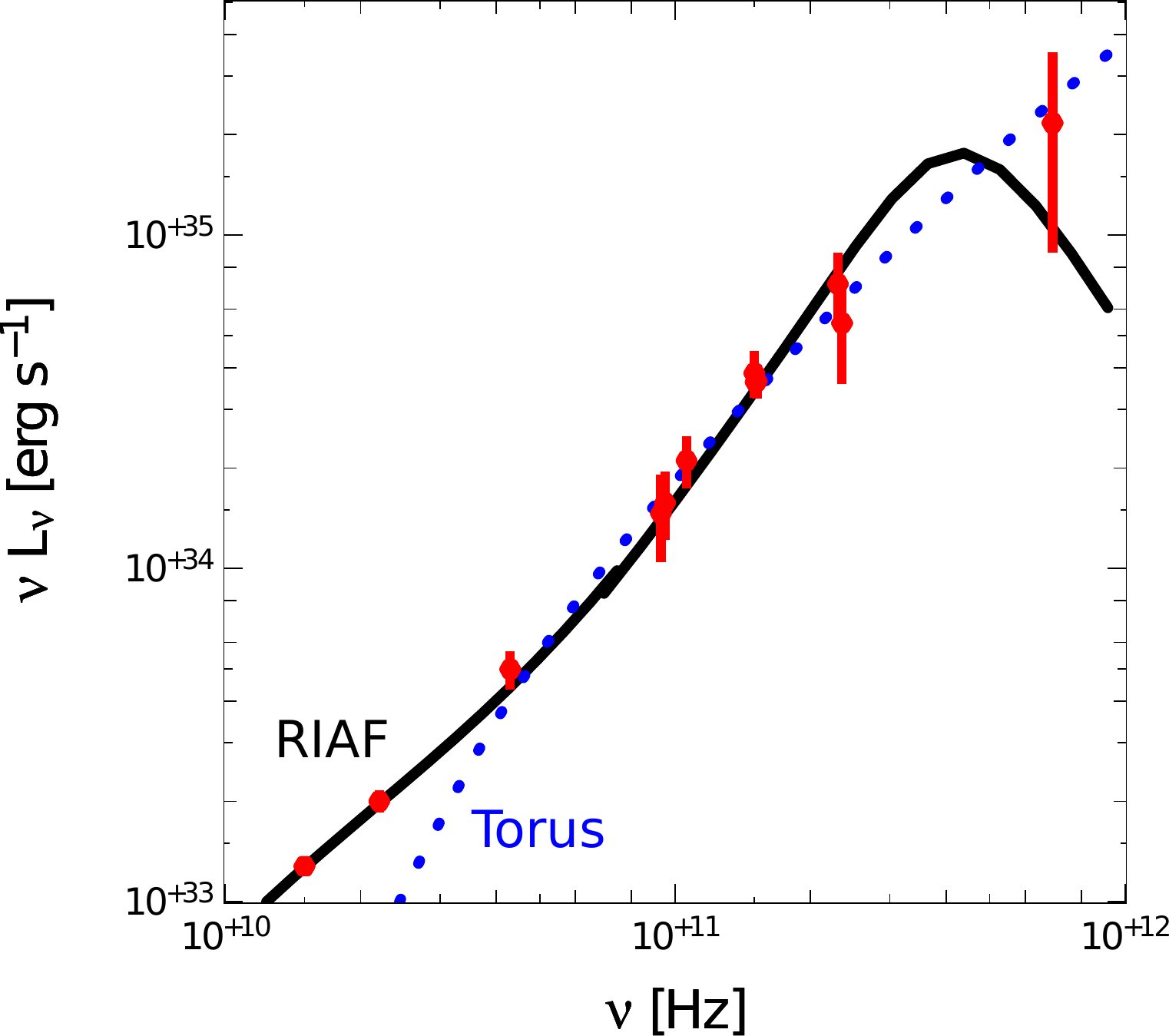}
\caption{Comparison of the best-fitting $a=0$, $i=5^\circ$ spectra as predicted
by the RIAF model (solid black) and the torus model (dotted blue). A non-thermal
power-law population of electrons is taken into account. The torus model is
not able to fit the slope of the centimeter data because of the compact distribution
of electrons, very different from the extended distribution of the RIAF model.} \label{fig:specPL}
\end{figure*}

Fig.~\ref{fig:specPL} shows the complete radio (centimeter to millimeter)
spectrum associated to $a=0$ and $i=5^\circ$, for both the torus
model and the RIAF model. We have modeled a RIAF following
\citet{broderick11}. We consider a power-law distribution for electron
density and temperature, following Eqs.~2 and~3 of \citet{broderick11}.
The non-thermal population is then constructed following the 
recipe given in Section~\ref{sec:PLelec}. Fig.~\ref{fig:specPL}
clearly shows that the RIAF model perfectly fits the whole radio
spectrum, while the torus model is not able to reproduce the 
correct centimeter slope. 

This difference between the two models is
due to the very different electron distribution functions. It is well
known that the $\nu F_\nu$ slope of power-law synchrotron
equals $7/2$. However, for an extended distribution of electrons,
different parts of the accretion flow will emit different spectra, 
and the averaging of these spectra can end up in a much shallower
slope. This is a well known effect in jets, and the same behavior
is found in the RIAF model. However, the torus model electron distribution
being much more compact, such an averaging is not obtained. We have
checked that even for very big tori, the non-thermal spectrum slope
is still too hard: indeed, as illustrated in Fig.~\ref{fig:nbdens}, even for
an extended torus, the electron distribution is clearly peaked towards
small distances. We have also checked that decreasing the outer
radius of the RIAF model leads to worse fits. For instance, in the
simulation used in Fig.~\ref{fig:specPL}, the outer radius of the RIAF
is at $r_{\mathrm{out}} = 300\,r_g$. Decreasing this value to 
$r_{\mathrm{out}} = 100\,r_g$ leads to a very bad fit, the centimeter
spectrum being too low (just as in the torus model), while the millimeter
data are still perfectly fit.

We conclude that the torus model is still unable to fit
the centimeter spectrum of Sgr~A*. Another possibility
to fit the centimeter data is to introduce a jet~\citep{falckemarkoff00}.
It is well known that Polish doughnuts, the base model of the
torus model, can naturally funnel jets~\citep{abramowicz78}. We plan in the future
to investigate whether a torus+jet analytic model will be
able to fit the whole radio specrum. The recent numeric
study of such a configuration by~\citet{moscibrodzka13} 
allows to be rather confident. In this perspective, the
recent searches for a putative jet at the Galactic center
are particularly meaningful~\citep{yusefzadeh12,li13}.

To conclude this Section, we highlight that our main result
as far as spectral data are concerned is to demonstrate that
the torus model is able to fit with perfect accuracy the
millimeter spectrum of Sgr~A*, and in particular the future
EHT band. However, much work remains to be done in order
to develop a complete synchrotron torus model of
Sgr~A*, from centimeter to near-infrared wavelengths.
Even though our model cannot yet be considered self-consistent,
we will investigate in the next Section the image properties
of the torus model at $1.3$~mm. At this wavelength, the
thermal synchrotron dominates, so we believe that our
purely thermal model will give realistic predictions. We
have checked that a pure thermal and a mixed thermal
power-law image for the RIAF model are very similar
at $1.3$~mm.

%-------------------------------------------------------------------
%-------------------------------------------------------------------
\section{Torus model images}
%-------------------------------------------------------------------
%-------------------------------------------------------------------

\subsection{$1.3$~mm images for best-fit spectral models}

Images of the best-fit models listed in Tab.~\ref{tab:parambf} are readily
computed at $1.3$~mm, a wavelength at which a stringent limit
on the size of the emitting region was imposed by \citet{doeleman08}
at $43^{+4.7}_{-2.7}\, \mu$as ($1\sigma$, scatter broadened).  We need to model
the smearing effect due to scattering by interstellar electrons. Following~\citet{moscibrodzka13}
we use a Gaussian profile with FWHM $1.309\,\lambda^2$~\citep{falcke00,bower06}.
At $1.3$~mm this boils down to smearing the image with a Gaussian of FWHM
$\approx 20\,\mu$as, already altering the image quite a lot.

\begin{figure*}[htbp]
\centering
\hspace{1.5cm}
\includegraphics[width=0.8\hsize]{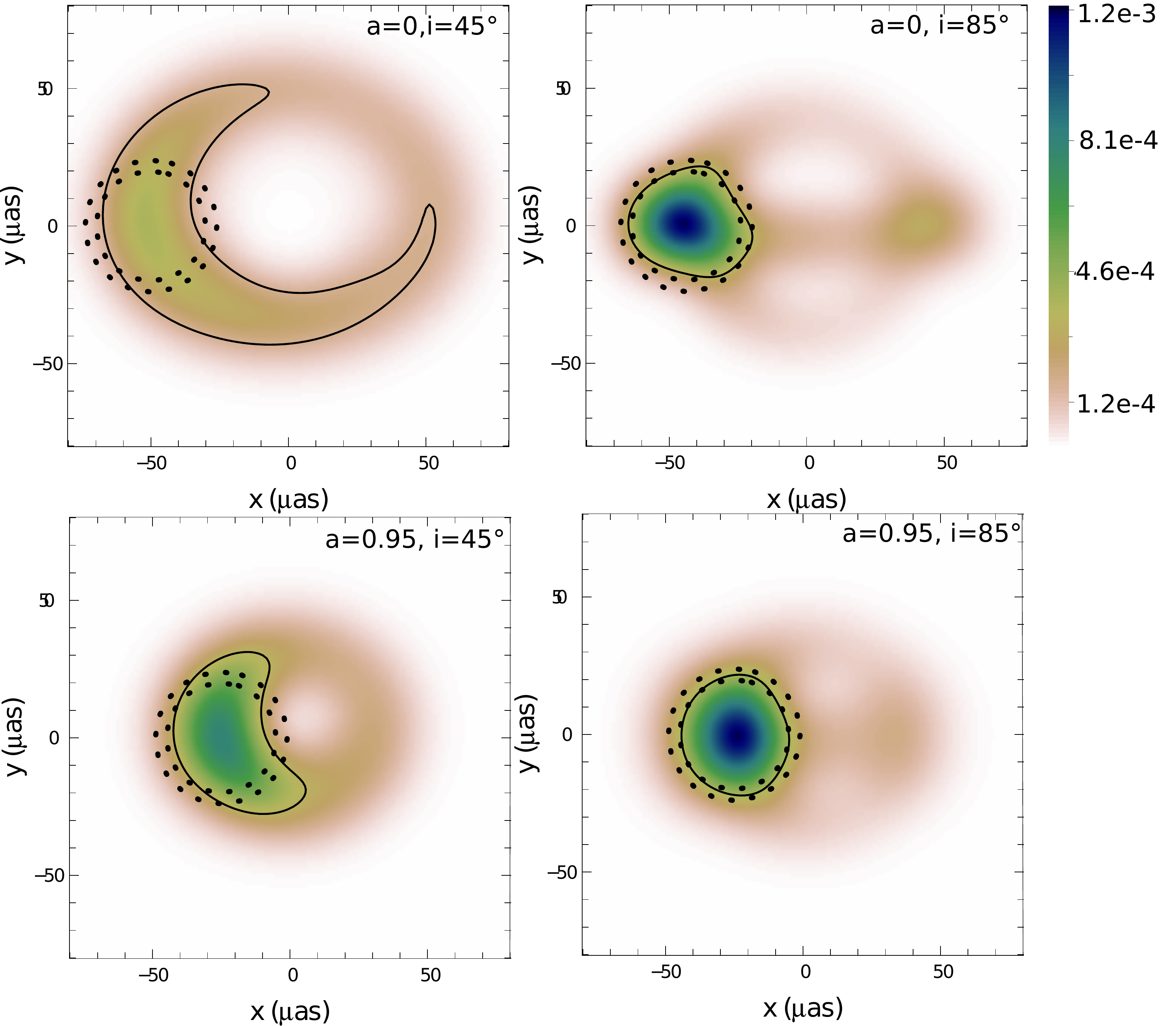}
\caption{Images (maps of specific intensity) at $1.3$~mm of the best-fit models of Tab.~\ref{tab:parambf}.
	The dotted circle show the $1\sigma$ confidence domain from \citet{doeleman08}.
	The thin solid curve encompass the region containing $50\%$ of the total flux.
	The color bar is common to all panels, and graduated in cgs units ($\mathrm{erg \, s^{-1}\,cm^{2}\,str^{-1}\,Hz^{-1}}$). } \label{fig:imagebf}
\end{figure*}

Fig.~\ref{fig:imagebf} shows the $1.3$~mm images corresponding to
the best-fit parameters of Tab.~\ref{tab:parambf}. Only the two higher
values of inclination were considered as the size limit imposed by
VLBI measurements favor higher inclination~\citep{broderick11}.
These images compare the size constraint of \citet{doeleman08}
to the size of the region emitting $50\%$ of the total flux. The figures
show that constraints on the inclination parameter are at hand: at low
spin, the $i=45^{\circ}$ case is clearly excluded, and marginally acceptable
for high spin. At $i=85^{\circ}$, the whole spin range will give a perfect
fit to the size constraint.

A complete study of the parameter space, fitting for the size of the
emitting region for a grid of spins and inclinations, goes beyond the
scope of the current article. It will be done in the future, once our
model will have been made self consistent over the whole radio
to infrared domain. The results presented here still support the
torus model as they show that the size predicted in high-inclination 
configuration fits well the VLBI size constraint at $1.3$~mm.

\subsection{Impact of magnetic field geometry on the image}

\begin{figure*}[htbp]
\centering
\includegraphics[width=0.8\hsize]{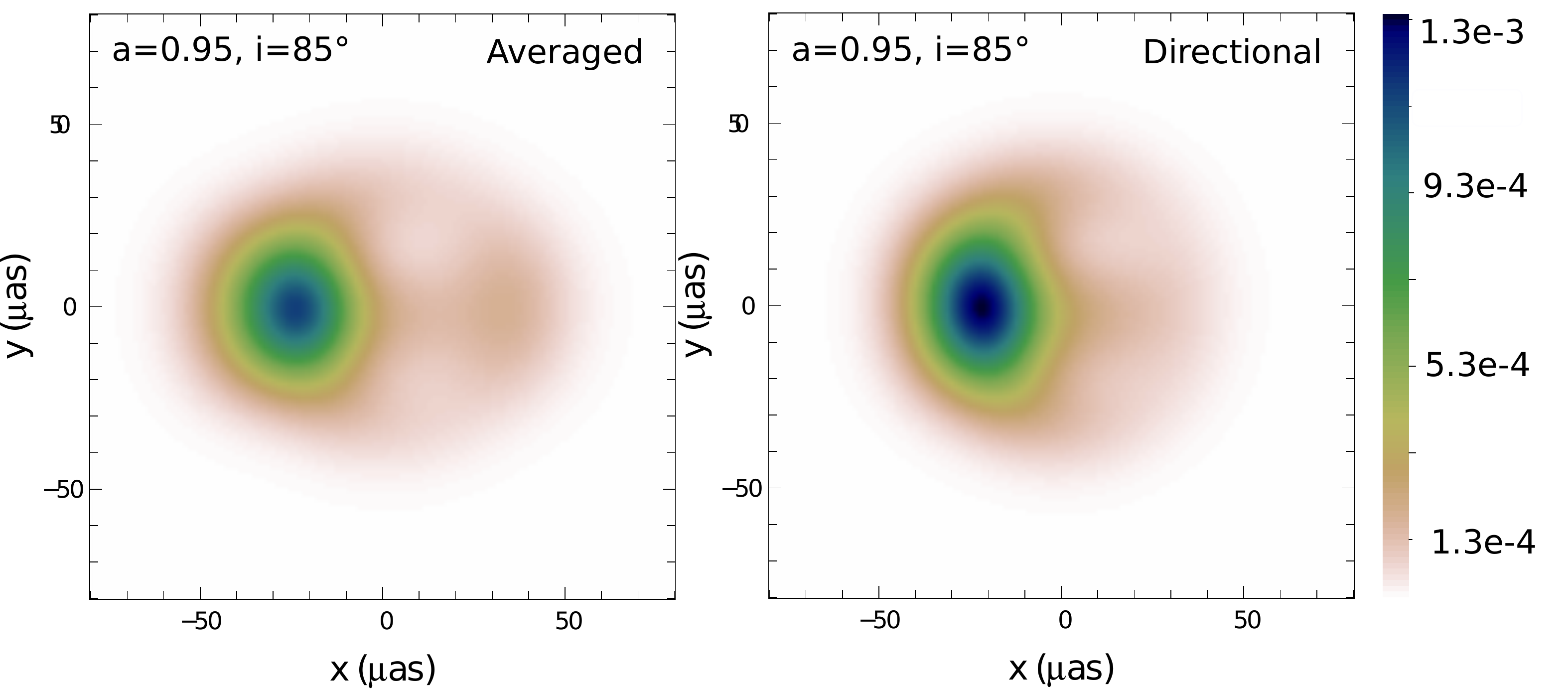}
\caption{Comparison of the best-fitting $a=0.95$, $i=85^\circ$ images 
at $1.3$~mm for an angle-averaged magnetic field (left) and a directional toroidal
magnetic field (right).} \label{fig:imageAvgDir}
\end{figure*}

Fig.~\ref{fig:imageAvgDir} shows the comparison of the
$1.3$~mm images associated to the best-fit spectral model at $a=0.95,
i=85^\circ$, for the angle-averaged and directional models.
It shows that the difference in the image is very weak, in the
same way as for the spectral data. It is thus very unlikely that
a constraint on the magnetic field geometry will be available by
means of spectral and imaging data. The most natural way to
get access to this information is to study the polarization
properties of the radiation~\citep[see e.g.][]{zamaninasab10}.

%---------------------------------------------------------------------
%---------------------------------------------------------------------
\section{Conclusion and perspectives}
%---------------------------------------------------------------------
%---------------------------------------------------------------------

We have constructed a millimeter-wavelength synchrotron radiative model for Sgr~A* based
on the fully general relativistic, analytical magnetized torus model of~\citet{komissarov06}. We
have considered the original model with a toroidal magnetic field, as well as an angle-averaged, 
chaotic-magnetic-field configuration. 
We show that such a model is able to account for the observable spectral constraints in the
millimeter domain. However, the torus model is not yet able of reproducing the centimeter
data. Constraints on the emitting region size at $1.3$~mm by \citet{doeleman08} are satisfied by the
torus model provided inclination is high enough. We also show that the magnetic field geometry
(chaotic or toroidal) cannot be constrained by spectral and imaging data.
This work shows that the torus model is a serious candidate to account for the properties
of Sgr~A* close environment in the millimeter domain, which in particular encompasses
the EHT observation band. 

Our torus model is the third analytic model proposed to account for Sgr~A* properties.
The jet model~\citep{falckemarkoff00} and the RIAF model~\citep{narayan95b,ozel00,yuan03,broderick11}
are alternatives. We have compared in this article the spectral predictions of the torus
and RIAF models. This comparison shows that the RIAF model is able to account
for the centimeter spectral data of Sgr~A* because of its extended distribution of
non-thermal electrons. As the torus model is firmly constrained to be compact
(within $\approx 15\,r_g$ from the black hole), it cannot produce the same spectrum.
Our goal is to develop in the near future a torus+jet model that would be able to
account for both spectral and imaging constraints over the radio band.

The development of such a self-consistent analytic model over the full synchrotron band
is interesting in the perspective of the future EHT data. Such an analytic model allows very fast computations as opposed
to GRMHD simulations. For example, one spectrum or one image at a relatively low resolution, 
sufficient to fit the data, requires a few minutes of computation on a standard laptop for
our model. This allows us to investigate large domains of physical parameters, which is
not doable with GRMHD simulations because of the computational time limitation.
In this perspective, we believe that the torus model for Sgr~A* as developed in this 
work will be of interest for the future data analysis linked with the EHT project. In particular,
this model may be a suitable test bed for investigating the observational counterparts of
compact objects alternative to the Kerr black hole.

\section*{Acknowledgements}
{The authors are particularly grateful to the referee, Heino Falcke, for his careful reading of their
first manuscript and his very accurate and relevant comments that allowed to greatly improve
the quality of this paper. 
FHV acknowledges interesting discussions with Monika Mo\'scibrodzka at the first \textit{Black Hole Cam} workshop in Effelsberg.
This work was supported by five Polish NCN grants: 2011/01/B/ST9/05439, 2012/04/M/ST9/00780, 2013/08/A/ST9/00795, 
2013/09/B/ST9/00060, and 2013/10/M/ST9/00729,
together with the European ``Synergy'' grant CZ.1.07/2.3.00/20.0071 aimed to support international collaboration at the In\-sti\-tu\-te of 
Physics of the Silesian University in Opava. 
This work was conducted within the scope of the HECOLS International Associated Laboratory, 
supported in part by the Polish NCN grant
DEC-2013/08/M/ST9/00664.
Computing was partly done using
Division Informatique de l'Observatoire (DIO) HPC facilities from
Observatoire de Paris (\url{http://dio.obspm.fr/Calcul/}).
}
%---------------------------------------------------------------------
%---------------------------------------------------------------------
%
%
%---------------------------------------------------------------------
%---------------------------------------------------------------------
\bibliography{bibkomis}

\begin{thebibliography}{37}
\expandafter\ifx\csname natexlab\endcsname\relax\def\natexlab#1{#1}\fi

\bibitem[{{Abramowicz} {et~al.}(1978){Abramowicz}, {Jaroszynski}, \&
  {Sikora}}]{abramowicz78}
{Abramowicz}, M., {Jaroszynski}, M., \& {Sikora}, M. 1978, \aap, 63, 221

\bibitem[{{Balick} \& {Brown}(1974)}]{balick74}
{Balick}, B. \& {Brown}, R.~L. 1974, \apj, 194, 265

\bibitem[{{Bower} {et~al.}(2006){Bower}, {Goss}, {Falcke}, {Backer}, \&
  {Lithwick}}]{bower06}
{Bower}, G.~C., {Goss}, W.~M., {Falcke}, H., {Backer}, D.~C., \& {Lithwick}, Y.
  2006, \apjl, 648, L127

\bibitem[{{Broderick} {et~al.}(2011){Broderick}, {Fish}, {Doeleman}, \&
  {Loeb}}]{broderick11}
{Broderick}, A.~E., {Fish}, V.~L., {Doeleman}, S.~S., \& {Loeb}, A. 2011, ApJ,
  735, 110

\bibitem[{{Broderick} {et~al.}(2014){Broderick}, {Johannsen}, {Loeb}, \&
  {Psaltis}}]{broderick14}
{Broderick}, A.~E., {Johannsen}, T., {Loeb}, A., \& {Psaltis}, D. 2014, \apj,
  784, 7

\bibitem[{{Broderick} \& {Loeb}(2006)}]{broderick06}
{Broderick}, A.~E. \& {Loeb}, A. 2006, ApJL, 636, L109

\bibitem[{{Chan} {et~al.}(2009){Chan}, {Liu}, {Fryer}, {Psaltis}, {{\"O}zel},
  {Rockefeller}, \& {Melia}}]{chan09}
{Chan}, C.-k., {Liu}, S., {Fryer}, C.~L., {et~al.} 2009, ApJ, 701, 521

\bibitem[{{Dexter} {et~al.}(2010){Dexter}, {Agol}, {Fragile}, \&
  {McKinney}}]{dexter10}
{Dexter}, J., {Agol}, E., {Fragile}, P.~C., \& {McKinney}, J.~C. 2010, ApJ,
  717, 1092

\bibitem[{{Doeleman} {et~al.}(2009){Doeleman}, {Agol}, {Backer}, {Baganoff},
  {Bower}, {Broderick}, {Fabian}, {Fish}, {Gammie}, {Ho}, {Honman},
  {Krichbaum}, {Loeb}, {Marrone}, {Reid}, {Rogers}, {Shapiro}, {Strittmatter},
  {Tilanus}, {Weintroub}, {Whitney}, {Wright}, \& {Ziurys}}]{doeleman09}
{Doeleman}, S., {Agol}, E., {Backer}, D., {et~al.} 2009, in Astronomy, Vol.
  2010, astro2010: The Astronomy and Astrophysics Decadal Survey, 68

\bibitem[{{Doeleman} {et~al.}(2008){Doeleman}, {Weintroub}, {Rogers},
  {Plambeck}, {Freund}, {Tilanus}, {Friberg}, {Ziurys}, {Moran}, {Corey},
  {Young}, {Smythe}, {Titus}, {Marrone}, {Cappallo}, {Bock}, {Bower},
  {Chamberlin}, {Davis}, {Krichbaum}, {Lamb}, {Maness}, {Niell}, {Roy},
  {Strittmatter}, {Werthimer}, {Whitney}, \& {Woody}}]{doeleman08}
{Doeleman}, S.~S., {Weintroub}, J., {Rogers}, A.~E.~E., {et~al.} 2008, \nat,
  455, 78

\bibitem[{{Falcke} \& {Markoff}(2000)}]{falckemarkoff00}
{Falcke}, H. \& {Markoff}, S. 2000, A\&A, 362, 113

\bibitem[{{Falcke} {et~al.}(2000){Falcke}, {Melia}, \& {Agol}}]{falcke00}
{Falcke}, H., {Melia}, F., \& {Agol}, E. 2000, \apjl, 528, L13

\bibitem[{{Genzel} {et~al.}(2010){Genzel}, {Eisenhauer}, \&
  {Gillessen}}]{genzel10}
{Genzel}, R., {Eisenhauer}, F., \& {Gillessen}, S. 2010, Reviews of Modern
  Physics, 82, 3121

\bibitem[{{Ghez} {et~al.}(2008){Ghez}, {Salim}, {Weinberg}, {Lu}, {Do}, {Dunn},
  {Matthews}, {Morris}, {Yelda}, {Becklin}, {Kremenek}, {Milosavljevic}, \&
  {Naiman}}]{ghez08}
{Ghez}, A.~M., {Salim}, S., {Weinberg}, N.~N., {et~al.} 2008, \apj, 689, 1044

\bibitem[{{Gillessen} {et~al.}(2009{\natexlab{a}}){Gillessen}, {Eisenhauer},
  {Fritz}, {Bartko}, {Dodds-Eden}, {Pfuhl}, {Ott}, \& {Genzel}}]{gillessen09a}
{Gillessen}, S., {Eisenhauer}, F., {Fritz}, T.~K., {et~al.} 2009{\natexlab{a}},
  \apjl, 707, L114

\bibitem[{{Gillessen} {et~al.}(2009{\natexlab{b}}){Gillessen}, {Eisenhauer},
  {Trippe}, {Alexander}, {Genzel}, {Martins}, \& {Ott}}]{gillessen09b}
{Gillessen}, S., {Eisenhauer}, F., {Trippe}, S., {et~al.} 2009{\natexlab{b}},
  \apj, 692, 1075

\bibitem[{{Goldston} {et~al.}(2005){Goldston}, {Quataert}, \&
  {Igumenshchev}}]{goldston05}
{Goldston}, J.~E., {Quataert}, E., \& {Igumenshchev}, I.~V. 2005, ApJ, 621, 785

\bibitem[{{Komissarov}(2006)}]{komissarov06}
{Komissarov}, S.~S. 2006, \mnras, 368, 993

\bibitem[{{Leahy}(1991)}]{hughes91}
{Leahy}, J.~P. 1991, {in Hughes, P.~A., ed. , Beams and jets in astrophysics,
  Cambridge University Press, Cambridge (p. 100)}

\bibitem[{{Li} {et~al.}(2013){Li}, {Morris}, \& {Baganoff}}]{li13}
{Li}, Z., {Morris}, M.~R., \& {Baganoff}, F.~K. 2013, ApJ, 779, 154

\bibitem[{{Marrone} {et~al.}(2006){Marrone}, {Moran}, {Zhao}, \&
  {Rao}}]{marrone06}
{Marrone}, D.~P., {Moran}, J.~M., {Zhao}, J.-H., \& {Rao}, R. 2006, Journal of
  Physics Conference Series, 54, 354

\bibitem[{{McKinney} {et~al.}(2013){McKinney}, {Tchekhovskoy}, {Sadowski}, \&
  {Narayan}}]{mckinney13}
{McKinney}, J.~C., {Tchekhovskoy}, A., {Sadowski}, A., \& {Narayan}, R. 2013,
  ArXiv e-prints

\bibitem[{{Mo{\'s}cibrodzka} \& {Falcke}(2013)}]{moscibrodzka13}
{Mo{\'s}cibrodzka}, M. \& {Falcke}, H. 2013, \aap, 559, L3

\bibitem[{{Mo{\'s}cibrodzka} {et~al.}(2009){Mo{\'s}cibrodzka}, {Gammie},
  {Dolence}, {Shiokawa}, \& {Leung}}]{moscibrodzka09}
{Mo{\'s}cibrodzka}, M., {Gammie}, C.~F., {Dolence}, J.~C., {Shiokawa}, H., \&
  {Leung}, P.~K. 2009, ApJ, 706, 497

\bibitem[{{Narayan} \& {Yi}(1995)}]{narayan95}
{Narayan}, R. \& {Yi}, I. 1995, \apj, 452, 710

\bibitem[{{Narayan} {et~al.}(1995){Narayan}, {Yi}, \& {Mahadevan}}]{narayan95b}
{Narayan}, R., {Yi}, I., \& {Mahadevan}, R. 1995, Nature, 374, 623

\bibitem[{{Noble} {et~al.}(2007){Noble}, {Leung}, {Gammie}, \&
  {Book}}]{noble07}
{Noble}, S.~C., {Leung}, P.~K., {Gammie}, C.~F., \& {Book}, L.~G. 2007,
  Classical and Quantum Gravity, 24, 259

\bibitem[{{{\"O}zel} {et~al.}(2000){{\"O}zel}, {Psaltis}, \&
  {Narayan}}]{ozel00}
{{\"O}zel}, F., {Psaltis}, D., \& {Narayan}, R. 2000, ApJ, 541, 234

\bibitem[{{Petrosian} \& {McTiernan}(1983)}]{petrosian83}
{Petrosian}, V. \& {McTiernan}, J.~M. 1983, Phys. of Fluids, 3023, 26

\bibitem[{{Shcherbakov} {et~al.}(2012){Shcherbakov}, {Penna}, \&
  {McKinney}}]{shcher12}
{Shcherbakov}, R.~V., {Penna}, R.~F., \& {McKinney}, J.~C. 2012, ApJ, 755, 133

\bibitem[{{Straub} {et~al.}(2012){Straub}, {Vincent}, {Abramowicz},
  {Gourgoulhon}, \& {Paumard}}]{straub12}
{Straub}, O., {Vincent}, F.~H., {Abramowicz}, M.~A., {Gourgoulhon}, E., \&
  {Paumard}, T. 2012, \aap, 543, A83

\bibitem[{{Vincent} {et~al.}(2011){Vincent}, {Paumard}, {Gourgoulhon}, \&
  {Perrin}}]{vincent11}
{Vincent}, F.~H., {Paumard}, T., {Gourgoulhon}, E., \& {Perrin}, G. 2011,
  Classical and Quantum Gravity, 28, 225011

\bibitem[{{Wardzi{\'n}ski} \& {Zdziarski}(2000)}]{wardzinski00}
{Wardzi{\'n}ski}, G. \& {Zdziarski}, A.~A. 2000, \mnras, 314, 183

\bibitem[{{Yuan} \& {Narayan}(2014)}]{yuan14}
{Yuan}, F. \& {Narayan}, R. 2014, ArXiv e-prints: 1401.0586

\bibitem[{{Yuan} {et~al.}(2003){Yuan}, {Quataert}, \& {Narayan}}]{yuan03}
{Yuan}, F., {Quataert}, E., \& {Narayan}, R. 2003, \apj, 598, 301

\bibitem[{{Yusef-Zadeh} {et~al.}(2012){Yusef-Zadeh}, {Arendt}, {Bushouse},
  {Cotton}, {Haggard}, {Pound}, {Roberts}, {Royster}, \&
  {Wardle}}]{yusefzadeh12}
{Yusef-Zadeh}, F., {Arendt}, R., {Bushouse}, H., {et~al.} 2012, ApJL, 758, L11

\bibitem[{{Zamaninasab} {et~al.}(2010){Zamaninasab}, {Eckart}, {Witzel},
  {Dovciak}, {Karas}, {Sch{\"o}del}, {Gie{\ss}{\"u}bel}, {Bremer},
  {Garc{\'{\i}}a-Mar{\'{\i}}n}, {Kunneriath}, {Mu{\v z}i{\'c}}, {Nishiyama},
  {Sabha}, {Straubmeier}, \& {Zensus}}]{zamaninasab10}
{Zamaninasab}, M., {Eckart}, A., {Witzel}, G., {et~al.} 2010, A\&A, 510, A3

\end{thebibliography}
\bibliographystyle{aa}
%---------------------------------------------------------------------
%---------------------------------------------------------------------
\end{document}